\documentclass[twocolumn,american,prl,superscriptaddress]{revtex4}
\usepackage[T1]{fontenc}
\usepackage[latin1]{inputenc}
\usepackage{graphicx}
\usepackage{amssymb}

\makeatletter

\usepackage{ae,aecompl}

\usepackage{babel}
\makeatother
\begin{document}

\title{Vorticity cutoff in nonlinear photonic crystals}

\author{Albert Ferrando}

\affiliation{Departament d'Òptica, Universitat de València. Dr. Moliner, 50. E-46100
Burjassot (València), Spain.}

\affiliation{Departamento de Matemática Aplicada, Universidad Politécnica de Valencia.
Camino de Vera, s/n. E-46022 Valencia, Spain.}

\author{Mario Zacarés}

\affiliation{Departament d'Òptica, Universitat de València. Dr. Moliner, 50. E-46100
Burjassot (València), Spain.}

\author{Miguel-Ángel García-March }

\affiliation{Departament d'Òptica, Universitat de València. Dr. Moliner, 50. E-46100
Burjassot (València), Spain.}

\date{\today}

\begin{abstract}
Using group theory arguments, we demonstrate that, unlike in homogeneous
media, no symmetric vortices of arbitrary order can be generated in
two-dimensional (2D) nonlinear systems possessing a discrete-point
symmetry. The only condition needed is that the non-linearity term
exclusively depends on the modulus of the field. In the particular
case of 2D periodic systems, such as nonlinear photonic crystals or
Bose-Einstein condensates in periodic potentials, it is shown that
the realization of discrete symmetry forbids the existence of symmetric
vortex solutions with vorticity higher than two. 
\end{abstract}


\maketitle

Vortices are particular higher-order stationary solutions present
in many different nonlinear systems, ranging from fluid dynamics to
photonics. A vortex is characterized by a typical phase dislocation
determined by an integer number, that we refer to as vorticity (also
known as winding-number, {}``topological charge'' or even spin).
An optical vortex with a rotationally invariant amplitude in a nonlinear
Kerr medium, experimentally observed in homogeneous self-defocussing
media \cite{swartzlander-prl69_2503}, can be understood as an eigenmode
of the equivalent rotationally invariant waveguide generated by itself
\cite{snyder-ol17_789}. Thus, a vortex appear as an object carrying
well-defined angular momentum: $\phi_{l}=e^{il\theta}f(r)$. In this
case, angular momentum and vorticity are the same integer number;
a consequence of the continuous $O(2)$-symmetry of the operator defining
the equivalent waveguide. However, in systems such as 2D nonlinear
photonic crystals or Bose-Einstein condensates in 2D periodic traps
this $O(2)$-symmetry is replaced by a discrete point-symmetry. Angular
momentum is no longer well defined and thus the angular-momentum-vorticity
equivalence is lost. Nevertheless, optical vortices have been predicted
to exist in 2D periodic photonic crystals \cite{yang-ol28_2094,alexander-prl93_63901}
and in photonic crystal fibers \cite{ferrando-oe12_817} and experimentally
observed in optically-induced photonic lattices \cite{neshev-prl92_123903}.
Although these solutions cannot longer have well-defined angular momentum,
certainly all of them present neat phase dislocations that can be
characterized by an integer vorticity value. In this paper, we will
prove how to re-interpret vorticity in terms of the rotational properties
of vortex solutions without resorting to the angular-momentum concept.
As a result, severe restrictions on vorticity values will be found
using group-theory arguments.

Let us consider the following general nonlinear equation for stationary
states:\begin{equation}
\left[L_{0}+L_{\mathrm{NL}}(|\phi|)\right]\phi(x,y)=-\mathcal{E}\phi(x,y),\label{eq:stationary_states}\end{equation}
where $L_{0}$ is a linear field-independent self-adjoint operator
(normally dependent on gradients and functions of the transverse coordinates)
and $L_{\mathrm{NL}}(|\phi|)$ is the nonlinear field-dependent piece
of the full operator acting on the field $\phi$. This equation is
valid for all type of 2D systems in which the nonlinearity depends
on the field through its modulus. Many different systems can be modeled
using an equation that can be written in the form given by Eq.(\ref{eq:stationary_states}).
We are interested in systems that, besides being described by Eq.(\ref{eq:stationary_states}),
are invariant under some discrete-symmetry group $G$: $[L,G]=0$
($L\equiv L_{0}+L_{\mathrm{NL}}$). This means that we assume that
all linear and nonlinear coefficients appearing in the operators defining
Eq.(\ref{eq:stationary_states}) are invariant under the $G$ group.
Our goal is to study the implications that the realization of discrete
symmetry have on the characterization of vortex solutions of Eq.(\ref{eq:stationary_states}).

The key concept in our approach is the so-called group self-consistency
condition. This condition establishes that if a system described by
Eq.(\ref{eq:stationary_states}) is invariant under some discrete-symmetry
group $G$ then any of its solutions either belongs to one representation
of the group $G$ or to one of its subgroups $G'$ ($G'\subset G$)\cite{ferrando-arXiv:nlin_0409045}.
Note that the identity group is always a subgroup of any group and,
therefore, asymmetric solutions also satisfy the group self-consistency
condition \cite{alexander-prl93_63901}. In this letter, however,
we will focus on symmetric solutions exclusively.

\begin{figure}
\includegraphics{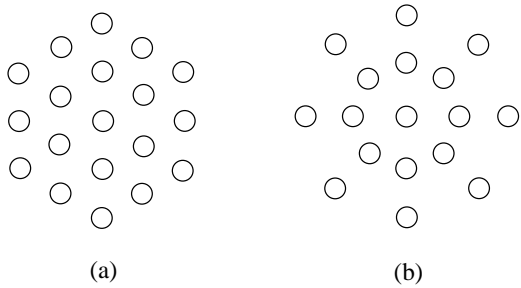}

\caption{Two examples of structures invariant under $2\pi/n$ rotations plus
specular reflections on the $x$ and $y$ axis: (a) 6-fold rotation
axis ($\mathcal{C}_{6v}$ group), and (b) 8-fold rotation axis ($\mathcal{C}_{8v}$
group).\label{cap:Two-examples-of-C6v-and-C8v}}
\end{figure}

The elements of point-symmetry groups in a plane are rotations through
integral multiples of $2\pi/n$ about some axis (called an $n$-fold
rotation axis), reflections on a mirror plane containing the axis
of rotation and combinations of both. Groups containing a $n$-fold
rotation axis constitute the \emph{$\mathcal{C}_{n}$} groups. When,
in combination with the $n$-fold rotation axis, these groups have
mirror planes, one generates the so-called $\mathcal{C}_{nv}$ groups.
In Fig. \ref{cap:Two-examples-of-C6v-and-C8v} we give two examples
of structures exhibiting $\mathcal{C}_{6v}$ and $\mathcal{C}_{8v}$
point-symmetries.

\begin{figure}
\includegraphics{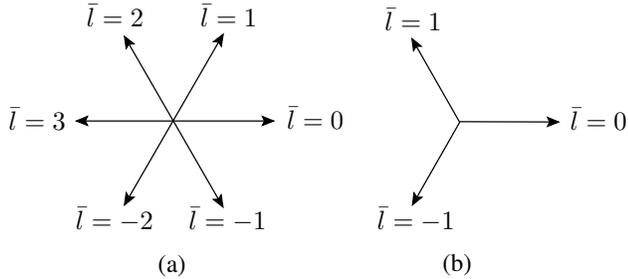}

\caption{Roots of unity diagrams displaying the representations of: (a) $\mathcal{C}_{6}$
and (b) $\mathcal{C}_{3}$.\label{cap:Roots-of-unity}}
\end{figure}

 We will prove next how vorticity is affected by the finite order
of the $n$-fold rotation axis defining the $\mathcal{C}_{n}$ (or
$\mathcal{C}_{nv}$) group. In order to do so, we need first to properly
characterize the different representations of a $\mathcal{C}_{n}$
group. Since $\mathcal{C}_{n}$ groups are abelian, its representations
are one-dimensional and given by a single scalar complex number (the
character of the representation) \cite{hamermesh64}. This scalar
is nothing but a root of unity of order $n$ and thus the representations
of the $\mathcal{C}_{n}$ group are given by $\{1,\epsilon^{\pm1},\dots,\epsilon^{\pm l},\dots,\epsilon^{n/2}\}$
for even $n$ and $\{1,\epsilon^{\pm1},\dots,\epsilon^{\pm l},\dots,\epsilon^{\pm(n-1)/2}\}$
for odd $n$, where $\epsilon=\exp(2\pi i/n)$. In Fig. \ref{cap:Roots-of-unity}
we present, as an example, the construction of the roots of unity
for the $\mathcal{C}_{6}$ (even $n$) and $\mathcal{C}_{3}$ (odd
$n$) groups. Each representation can be labeled by the natural number
$l$ and, when present, by its sign. We denote it by $\mathcal{D}_{l,s}$
($l\in\mathbb{N}$, $s=\pm$). No sign is needed for the identity
representation $\mathcal{D}_{0}$ ($l=0$) nor for $\mathcal{D}_{n/2}$
($l=n/2$, even $n$). A state belonging to representations with $l\neq0,n/2$
can be written as $\left|l,s\right\rangle $ with $0<l<n/2$ (if $n$
is even) or $0<l\le(n-1)/2$ (if $n$ is odd). When we act with a
group operator $G$ (representing a discrete rotation of angle $2\pi/n$)
on a function belonging to a representation $\mathcal{D}_{l,s}$,
it transforms as $G\phi_{\bar{l}}=\epsilon^{\bar{l}}\phi_{\bar{l}}$,
where $\bar{l}\equiv sl$ ($s=\pm,$$l\in\mathbb{N}$). Clearly too,
$G\phi_{0}=\phi_{0}$ and $G\phi_{n/2}=\epsilon^{n/2}\phi_{n/2}$
(even $n$). If there is no other symmetry involved, $[L,G]=0$ implies
that every one-dimensional representation is characterized by a different
$L$-eigenvalue: $L\phi_{\bar{l}}=-\mathcal{E}_{\bar{l}}\phi_{\bar{l}}$.
Representations are thus non-degenerated.

We proceed now to explicitly construct functions belonging to $l\neq0$
representations. Let us consider the complex coordinate vector $u=x+iy=re^{i\theta}$.
Integer powers of $u$ have well-defined transformation properties
under a $2\pi/n$ rotation: $u^{\bar{l}}\stackrel{\theta\rightarrow\theta+2\pi/n}{\rightarrow}\epsilon^{\bar{l}}u^{\bar{l}}$.
Therefore, we can easily construct a function in the $\mathcal{D}_{l,s}$
representation of $\mathcal{C}_{n}$ as \begin{equation}
\phi_{\bar{l}}(u)=u^{\bar{l}}\phi_{0}^{(\bar{l})}(u),\label{eq:l-representation}\end{equation}
 $\phi_{0}^{(\bar{l})}$ being a function in the $\mathcal{D}_{0}$
representation of $\mathcal{C}_{n}$. Clearly, $G\phi_{\bar{l}}=\epsilon^{\bar{l}}\phi_{\bar{l}}$.

\begin{figure}
\includegraphics{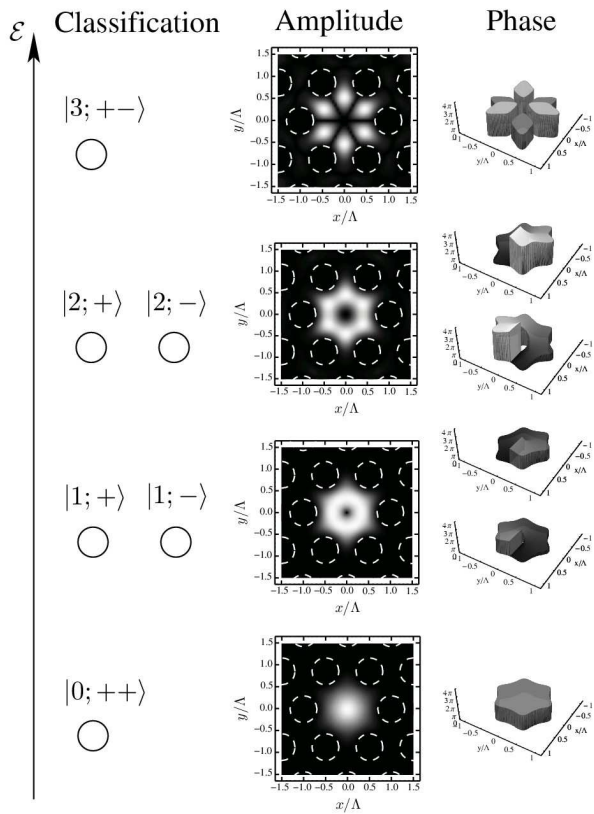}

\caption{Lowest order eigenfunctions of a nonlinear operator $L$ generated
by a soliton solution in the identity (fundamental) representation
of $\mathcal{C}_{6v}$. The symmetry of the full operator is $\mathcal{C}_{6v}$:
$[L,\mathcal{C}_{6v}]=0$.\label{cap:Lowest-order-eigenfunctions}}
\end{figure}

The representations of $\mathcal{C}_{nv}$ (discrete rotations plus
reflections) are easily obtained from those of $\mathcal{C}_{n}$
groups \cite{hamermesh64}. The existence of the extra symmetries
provided by mirror reflections yields to degeneracies for $\bar{l}\neq0$
representations. High-order states are now doubly degenerated; they
form pairs of complex-conjugated functions ($\phi_{l}$,$\phi_{l}^{*}$)
with the same $L$-eigenvalue: $L\phi_{\bar{l}}=-\mathcal{E}_{l}\phi_{\bar{l}}$.
Remarkable exceptions are the $\mathcal{D}_{0}$ and $\mathcal{D}_{n/2}$
representations. Because of their different behavior under mirror
reflections there are two distinct non-degenerated one-dimensional
$l=0$ representations: $\left|0;++\right\rangle $ and $\left|0;--\right\rangle $.
They transform differently under reflections with respect to $x$
and $y$ axis: $R_{x,y}\left|0;++\right\rangle =+\left|0;++\right\rangle $
and $R_{x,y}\left|0;--\right\rangle =-\left|0;--\right\rangle $.
The $\left|0;++\right\rangle $ state has maximal symmetry. Fundamental
solitons belongs to this identity representation of $\mathcal{C}_{nv}$.
In the same way, there are also two different non-degenerated one-dimensional
representations with $l=n/2$ (even $n$): $\left|n/2;+-\right\rangle $
and $\left|n/2;-+\right\rangle $. The distinction is made by $R_{x,y}$-reflections.
In Fig. \ref{cap:Lowest-order-eigenfunctions} we show the lowest
order eigenfunctions of the spectrum of a $\mathcal{C}_{6v}$-invariant
operator self-consistently generated by a fundamental soliton solution
($\phi_{\mathrm{fund}}=\phi_{0,++}$): $L=L_{0}+L_{\mathrm{NL}}(\left|\phi_{\mathrm{fund}}\right|)$
(see the final section of the paper for details on the physical system
associated to $L$). We easily recognize, from lower to higher values
of $\mathcal{E}$, the $\left|0;++\right\rangle $ self-consistent
state (i.e., the fundamental soliton), the doubly-degenerated $\left|1;\pm\right\rangle $
and $\left|2;\pm\right\rangle $ states and the non-degenerated $\left|3;+-\right\rangle $
state. The rest of the spectrum, including continuum de-localized
states, systematically falls into the representations described above.

Vorticity $v$ can be defined as the integer variation (in 2$\pi$
units) that the phase of a complex field experiments under a $2\pi$
rotation around a rotation axis. Solutions with non-zero vorticity
are called vortices of order $v$. They are characterized by their
rotation axis, whose intersection with the 2D plane defines the vortex
center, where their amplitude vanishes. If $\Phi(r,\theta)$ represents
the phase of a complex vortex field of order $v$ given by $f_{v}=|f_{v}|e^{i\Phi}$,
then $\Phi(r,\theta+2\pi)-\Phi(r,\theta)=2\pi v$, where the polar
coordinates are referred to a reference frame centered on the rotation
axis. For systems enjoying a 2D point symmetry, this axis is naturally
given by the $n$-fold rotation axis of the corresponding $\mathcal{C}_{n}$
(or $\mathcal{C}_{nv}$) group. 

According to the group self-consistency condition, all symmetric solutions
of Eq.(\ref{eq:stationary_states}) in a system with $\mathcal{C}_{n}$
symmetry have to lie on the representations of $\mathcal{C}_{n}$
or of any of its subgroups. Let us consider now a solution $\phi_{\bar{l}}$
in the $\mathcal{D}_{l,s}$ representation of $\mathcal{C}_{n}$ given
by Eq.(\ref{eq:l-representation}). Its phase will be given by $\arg\phi_{\bar{l}}(r,\theta)=\bar{l}\theta+\arg\phi_{0}^{(\bar{l})}(r,\theta)$.
Since $\phi_{0}^{(\bar{l})}(r,\theta)$ is invariant under rotations,
$\arg\phi_{\bar{l}}(r,\theta+2\pi)=\arg\phi_{\bar{l}}(r,\theta)+2\pi\bar{l}$.
Therefore we find the important relation between the index representation
and vorticity:\begin{equation}
v=\bar{l}.\label{eq:vorticity}\end{equation}
Vortices are thus solutions belonging to $\mathcal{D}_{l,s}$ representations
with $l\neq0$. There is, however, no vortex associated to $l=n/2$
(even $n$). It can be proved that $\phi_{n/2}$ is a real field,
so that its argument is a function that can only take the values $0$
or $\pi$. More explicitly, from Eq.(\ref{eq:l-representation}),
$\phi_{n/2}\sim\cos\left(n\theta/2+\arg\phi_{0}^{(n/2)}(r,\theta)\right)$,
which has the phase behavior of alternating signs typical of a nodal
soliton and not of a vortex \cite{ferrando-arXiv:nlin_0409045}. In
$\mathcal{C}_{nv}$, the behavior of the $\phi_{n/2,+-}$ and $\phi_{n/2,-+}$
functions is also of the nodal-soliton type, as one can check by observing
the phase of the $\left|3;+-\right\rangle $ state in Fig. \ref{cap:Lowest-order-eigenfunctions}. 

Let us summarize now our main conclusions. Firstly, if a system is
invariant under a $\mathcal{C}_{n}$ or $\mathcal{C}_{nv}$ point-symmetry
group, the solutions of Eq.(\ref{eq:stationary_states}) belong to
representations of these groups or of their corresponding subgroups.
Secondly, symmetric solutions of Eq.(\ref{eq:stationary_states})
are characterized by the representation index $l$, which has an upper
bound fixed by the order of the group: $l\le n/2$ (even $n$) and
$l\leq(n-1)/2$ (odd $n$). Thirdly, the vorticity $v$ of the vortex
solutions of such a system has a cutoff due to Eq. (\ref{eq:vorticity})
and the upper bound for $l$:\begin{equation}
\left|v\right|<n/2\,\,\,(\mathrm{even}\,\, n)\,\,\,\mathrm{and}\,\,\,\left|v\right|\le(n-1)/2\,\,\,(\mathrm{odd}\,\, n).\label{eq:cutoff}\end{equation}
Note that the group of continuous rotations on a plane can be understood
as the limiting case $O(2)=\lim_{n\rightarrow\infty}\mathcal{C}_{n}$
and, thus, Eq.(\ref{eq:cutoff}) correctly establishes the absence
of a cutoff for it ($\left|v\right|<\infty$).

When we deal with 2D periodic systems, the realization of discrete
symmetry has particular features. This is a well-known problem in
crystallography \cite{ashcroft76}. The crystal structure is constructed
according to a pattern that repeats itself to {}``tessellate'' the
2D plane. Patterns, unlike objects, are invariant under translations
(defined by the periodicity of the crystal). This translation property,
inherent to periodicity, determines that only certain collections
of symmetry elements are possible for patterns. In other words, only
patterns that exhibit a selected set of symmetries can {}``tessellate''
the 2D plane. The important result for us here is that pattern periodicity
establishes a restriction on the order of discrete rotations allowed
in plane groups. Only $n$-fold rotations of order 2,3,4 and 6 are
permitted in a 2D periodic crystal \cite{ashcroft76}. 

The previous group analysis has important implications for 2D nonlinear
periodic systems. According to the group self-consistent condition,
if a periodic system described by Eq.(\ref{eq:stationary_states})
is invariant under a discrete group $G$, its stationary solutions
cannot belong to representations of groups with higher symmetry than
$G$. If $G'$ is the symmetry group of the solution, then $G'\subseteq G$.
On the other hand, as seen above, the maximum $n$-fold rotation symmetry
compatible with periodicity is a sixth-fold rotation, which means
that the maximum value for the order $n$ of $\mathcal{C}_{n}$ and
$\mathcal{C}_{nv}$ point-symmetry groups in 2D periodic systems is
$n=6$. Consequently, the point-symmetry group of a solution cannot
exceed this order: $n\le6$. Since vorticity is restricted by the
order of the point-symmetry group according to Eq. (\ref{eq:cutoff}),
we come up to the conclusion that in 2D nonlinear periodic systems
of the type described by Eq.(\ref{eq:stationary_states}) vorticity
has a strict bound:$\left|v\right|\le2.$ Putting this into words,
there are no vortices of order higher than two in 2D nonlinear periodic
systems described by Eq.(\ref{eq:stationary_states}).

\begin{figure}
\includegraphics{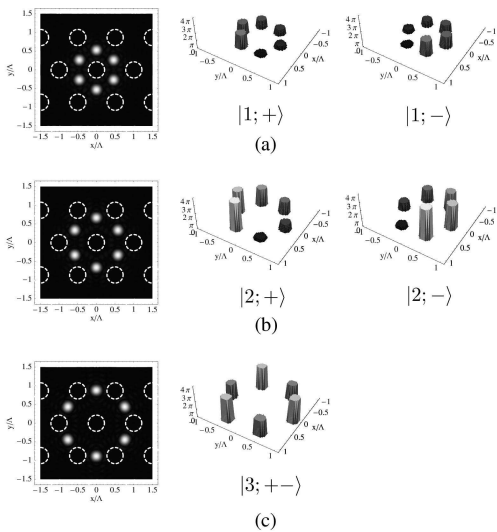}

\caption{Higher-order solitons for a periodic $\mathcal{C}_{6v}$ PCF: (a)-(b)
First- and second-order vortex pairs, $\left|1;\pm\right\rangle $
and $\left|2;\pm\right\rangle $; (c) nodal soliton of order three,
$\left|3;+-\right\rangle $.\label{cap:vortices_in_periodic_PCF} }
\end{figure}

\begin{figure}
\includegraphics{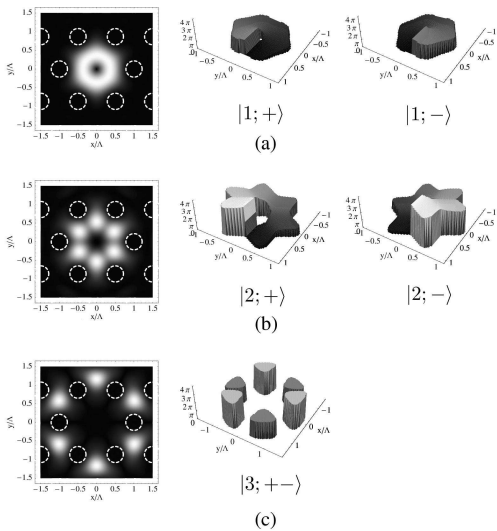}

\caption{Same as in Fig. (\ref{cap:vortices_in_periodic_PCF}) but in a PCF
with defect\label{cap:vortices_in_a_PCF_with_defect}.}
\end{figure}

In order to illustrate our previous theoretical results, we have numerically
studied a realistic system, namely, a photonic crystal fiber (PCF).
A PCF is a type of 2D photonic crystal consisting on a regular lattice
of holes in silica (characterized by the hole radius $a$ and the
lattice period --or pitch-- $\Lambda$) extending along the entire
fiber length. When one considers that the silica response is nonlinear
(nonlinearity represented by the nonlinear coefficient $\gamma$,
defined in Ref.\cite{ferrando-oe12_817}), a PCF becomes a 2D nonlinear
photonic crystal. The nonlinear propagation modes of a PCF for monochromatic
illumination in the scalar approximation verify Eq.(\ref{eq:stationary_states})
with $\mathcal{E}=-\beta^{2}$, $\beta$ being the mode propagation
constant (see \cite{ferrando-oe12_817}). Among possible hole-distribution
geometries we choose that based on a triangular lattice with $\mathcal{C}_{6v}$
symmetry (see Fig. \ref{cap:Two-examples-of-C6v-and-C8v}(a)). The
reason of our symmetry choice is simple. As proved before, the $\mathcal{C}_{6v}$
group provides the highest vorticity solutions since it corresponds
to the maximal point-symmetry achievable in a 2D nonlinear photonic
crystal. In Fig. \ref{cap:vortices_in_periodic_PCF} we find the first
three (from lowest to highest value of $\beta^{2}$) higher-order
solitons of a perfectly periodic PCF (without defect) calculated for
the values $a=5\,\mu\mathrm{m}$, $\Lambda=26\,\mu\mathrm{m}$, and
$\lambda=1064\,\mathrm{nm}$ at $\gamma=0.01$. In Fig. \ref{cap:vortices_in_a_PCF_with_defect}
we present the same first three higher-order solitons but for a PCF
with periodicity broken by the presence of a defect (absence of a
hole). Note that in both cases the symmetry group is $\mathcal{C}_{6v}$
and that, in agreement with our previous result, the maximum vorticity
allowed is two. The soliton solution with $l=3$ is not a vortex.
As predicted by group theory, it presents a binary phase structure
(corresponding to a $\left|3;+-\right\rangle $ state) of the nodal-soliton
type \cite{ferrando-arXiv:nlin_0409045}. It is interesting to check
the generality and accuracy of the group-theory approach using these
numerical examples. The spectrum of higher-order soliton solutions
is perfectly explained by our previous group-theory arguments nevertheless
the periodic (Fig. \ref{cap:vortices_in_periodic_PCF}) and non-periodic
(Fig. \ref{cap:vortices_in_a_PCF_with_defect}) photonic crystal structures
present notable differences. Despite they share the same $\mathcal{C}_{6v}$
symmetry, a description in terms of weakly interacting localized fundamental
solitons on lattice sites (the equivalent of the tight-binding approximation
in solid state physics) \cite{alexander-prl93_63901} can only be
valid in the perfectly periodic case. As it is apparent in Fig. \ref{cap:vortices_in_periodic_PCF},
this localization feature is clear in the amplitude and phase of vortex
and nodal soliton solutions in the periodic PCF. However, single fundamental
solitons are no longer recognizable in the vortex and nodal solitons
of Fig. \ref{cap:vortices_in_a_PCF_with_defect} due to the presence
of the periodicity-breaking defect. One can think of a situation of
strongly interacting solitons causing the {}``tight-binding approximation''
to stop being valid. Despite this fact, our main results concerning
the nature of solutions and, more specifically, the restrictions on
vorticity remain valid with complete generality.

We are thankful to P. Fernández de Córdoba for useful discussions.
This work was financially supported by the Plan Nacional I+D+I (grant
TIC2002-04527-C02-02), MCyT (Spain) and FEDER funds. Authors also
acknowledge the financial support from the Generalitat Valenciana
(grants GV04B-390 and Grupos03/227). M. Z. gratefully acknowledges
Fundaci\'{o}n Ram\'{o}n Areces grant.

\bibliographystyle{/usr/src/latex/revtex4/apsrev.bst}
\bibliography{bib_general}

\end{document}